\documentclass{llncs}

\usepackage{makeidx}
\usepackage{graphicx}
\usepackage{subcaption}
\usepackage{enumerate}
\usepackage[noadjust]{cite}
\usepackage[hidelinks, bookmarks=false, draft]{hyperref}
\usepackage{multirow}
\usepackage{booktabs}
\usepackage{dblfloatfix}
\usepackage{marvosym}
\usepackage{enumitem}
\usepackage{hhline}

\begin{document}

\title{Is Task Board Customization Beneficial?}

\subtitle{An Eye Tracking Study}


\author{Oliver Karras \Letter \and Jil Kl\"under \and Kurt Schneider}


\institute{Software Engineering Group\\ Leibniz Universit\"at Hannover 
\\30167 Hannover, Germany\\
\email{Email:\{oliver.karras, jil.kluender, kurt.schneider\}@inf.uni-hannover.de}}


\maketitle

\begin{abstract}
The task board is an essential artifact in many agile development approaches. 
It provides a good overview of the project status. Teams often customize their 
task boards according to the team members' needs. They modify the structure 
of boards, define colored codings for different purposes, and introduce 
different card sizes. Although the customizations are intended to improve the 
task board's usability and effectiveness, they may also complicate its 
comprehension and use. The increased effort impedes the work of both the team 
and team externals. Hence, task board customization is in conflict with the 
agile practice of fast and easy overview for everyone.

In an eye tracking study with 30 participants, we compared an original task 
board design with three customized ones to investigate which design shortened 
the required time to identify a particular story card. Our findings yield that 
only the customized task board design with modified structures reduces the 
required time. The original task board design is more beneficial than 
individual colored codings and changed card sizes.

According to our findings, agile teams should rethink their current task board 
design. They may be better served by focusing on the original task board design 
and by applying only carefully selected adjustments. In case of customization, 
a task board's structure should be adjusted since this is the only beneficial 
kind of customization, that additionally complies more precisely with the 
concept of fast and easy project overview.

\end{abstract}

\section{Introduction}
Agile software development is a general term for a set of development approaches which focus on social aspects. These approaches aim at increasing the developers' productivity, delivering working software in time and minimizing the risk of failure within software projects \cite{Sharp.2009}. The core concept of agile development is based on fundamental values which are concretized by defined principles that are in turn fulfilled by certain practices \cite{Beck.2001}.

eXtreme programming (XP) \cite{Beck.2007} and Scrum \cite{Schwaber.2002} are the most commonly used and combined agile approaches \cite{Azizyan.2011, Pikkarainen.2008}.
One practice of XP is the \textit{informative workspace}. According to Beck and Andres, this practice is about how to ``make your workspace about your work. An interested observer should be able to walk into the team space and get a general idea of how the project is going within $15$ seconds'' \cite[p. 39f.]{Beck.2007}.
Cockburn \cite{Cockburn.2009} provides a similar concept of the so-called 
\textit{information radiator}. ``An information radiator displays information 
in a place where passerby can see it'' \cite[p. 114]{Cockburn.2009}. An 
information radiator has to fulfill two features -- representing information 
that changes over time, and requiring very little effort to view the display. 
In total, an implementation of these two concepts must be easy-to-use and offer 
a fast overview with minimal effort \cite{Rubart.2009}.

One implementation of both concepts is the \textit{task board} 
\cite{Sharp.2006}. It is one key artifact of agile development 
\cite{Azizyan.2011, Katsma.2013} which serves a dual purpose of supporting a 
team's work organization and constituting at a glance how much work is left 
\cite{Cohn.2012, Petre.2012}. Additionally, a task board allows communication 
and collaboration since it tracks and visualizes the software development 
process and thus simplifies its accessibility for everyone \cite{Perry.2008, 
Pikkarainen.2008}.

Although the original task board design of Cohn \cite{Cohn.2012} provides a 
clear overview, teams tend to customize their own \cite{Sharp.2008}. Sharp et 
al. \cite{Sharp.2009} analyzed six different mature XP teams and their task 
boards. They identified that the teams' task boards were consistent in terms of 
usage, but not regarding a particular design.
The different task board designs resulted from combinations of various 
customizations like modified structures, individual colored codings and changed 
card sizes. Customization itself is not serious since a task board can be 
easily and flexibly adjusted due to its physical nature \cite{Hajratwala.2012, 
Sharp.2008}. Additionally, agile approaches involve customization by offering 
corresponding degrees of freedom \cite{Pikkarainen.2008, Sutherland.2009}. 
Furthermore, any adjustment of a task board by an agile team according to its 
needs is plausible since the team members work with it every day 
\cite{Sharp.2008}.

However, multiple combined customizations complicated the maintenance and 
comprehensibility of a task board. In particular, the increased effort impedes 
the work of a team as well as team externals with a customized task board 
\cite{Hajratwala.2012, Liskin.2014, Sharp.2009}. Thus, the underlying practice 
of a task board as an \textit{informative workspace} for fast and easy project 
overview for everyone gets lost.

While the tight social and technical cohesion found in mature agile teams are 
not disputed, the effect of single practices like the informative 
workspace is little understood \cite{Sharp.2009}. Berczuk emphasizes that ``any 
team is best served by following the rules of the agile method with as few 
adjustments as possible'' \cite[p. 6]{Berczuk.2007}. Corresponding to 
Pikkarainen et al. \cite{Pikkarainen.2008}, adoption and change of agile 
practices are aspects of future studies. Therefore, we investigated whether 
specific single task board customizations contribute to a task board's usage 
in comparison with an original task board design. As an example, we focused on 
the identification of a particular story card as one main task of using a task 
board.

We conducted an eye tracking study to compare an original task board design 
corresponding to literature \cite{Cohn.2012, Perry.2008} with three customized 
ones. Each customized task board differs exactly in one single aspect from the 
original one, such as modified structures, individual colored codings, or 
changed card sizes. Each modification could contribute to achieving a better 
overview of a task board in order to identify a particular story card faster. 
In our study, we observe whether a particular task board customization improves 
the work with a task board. These results identify whether specific kinds of 
customization are beneficial or not for a task board's usage. Our findings can 
help agile teams to rethink their current task board design in order to improve 
it.

The contribution of this paper is the insight that modified structures are the 
only kind of customization that shortens time to identify a particular story 
card. Individual colored codings and changed card sizes even have detrimental 
effects on the performance. Agile teams should reconsider their current task 
board designs. They may be better served by focusing on the original task board 
design and applying carefully selected adjustments. A task board's structure 
should be adjusted since this kind of customization is beneficial and complies 
with the agile practice \textit{informative workspace}.

This paper is structured as follows: Section \ref{sec:related-work} discusses 
related work. We describe the task board and its major kinds of customization 
in section \ref{sec:task-board}. In section \ref{sec:controlled-experiment}, we 
report our eye tracking study and document its findings, which we discuss in 
section \ref{sec:discussion}. Section \ref{sec:conclusion} concludes the paper.

\section{Related Work}
\label{sec:related-work}

\subsection{Task Board: Key Artifact of Agile Software Development}
Several researchers investigated the task board's usage and role in the agile 
software development process.

Sharp et al. \cite{Sharp.2006} systematically consider the use and role of 
story cards and a task board in one mature XP team. Based on story cards and 
the task board, the authors analyze the team's collaborative work by using the 
distributed cognition framework. Thus, the information flows in, around and 
within the XP team can be substantiated to answer ``what if'' questions 
regarding changes to the story cards' and task board's form to illustrate 
consequences for the teamwork.
Sharp and Robinson \cite{Sharp.2008} extend the previously mentioned study on 
three mature XP teams. Their results show significant similarities between the 
teams' usage of story cards and task board, but not in their particular 
designs. After discussing the importance of a physical representation of both 
artifacts, the authors highlight important aspects that need to be taken into 
account for technological tool-support of agile development.
In a further study, Sharp et al. \cite{Sharp.2009} investigate the role of 
story cards and a task board from two complementary perspectives: a notational 
and a social one. Based on both perspectives, they explain that these two 
physical artifacts are important key properties of successful teams. Any 
attempt to replace these artifacts with technological support needs to take 
into account the complex relationships between both perspectives and the 
artifacts.
Petre et al. \cite{Petre.2012} consider the use of public visualizations, i.e. 
story cards and task boards, in different software development teams. In a 
number of empirical studies, the authors observe differences in the use of 
paper and whiteboards between traditional and agile teams. The findings are 
used to identify possible implications of these differences for software 
development in general.
Liskin et al. \cite{Liskin.2014} explore the use and role of story cards and 
task board within a Kanban project. Their findings reveal that despite a task 
board for requirements visualization and communication some requirements are 
still too implicit and caused misunderstandings.
Katsma et al. \cite{Katsma.2013} investigate the usage of software- and 
paper-based task boards in globally distributed agile development teams. They 
conclude that paper-based task boards currently offer many advantages compared 
to its software-based solutions. By applying the media synchronicity theory, 
Katsma et al. \cite{Katsma.2013} explain the current use and future development 
of software tools to support globally distributed agile development teams.
Perry \cite{Perry.2008} reports his experiences about transparency problems in 
agile teams due to difficulties in the transition from a physical to electronic 
task board. He discusses the advantages and disadvantages of physical and 
electronic task boards. Based on his observation, he concludes that both task 
board types have their place in team collaboration. However, the simple power 
and utility of physical task boards should not be neglected.
Hajratwala \cite{Hajratwala.2012} observes the creation and evolution of 
various task boards over time in different projects. He explains the reasons 
why the task boards evolved, and recommends key attributes that a task board 
should have.

The previous investigations focus on both the usage and role of story cards and 
task boards in agile software development. The main focus is on the general 
work with a task board and its importance for agile development. Additionally, 
different task board designs and their evolution over time are presented. 
Although differences in the designs were recognized, none of the researchers 
considered its possible impact on work with this artifact. Our paper addresses 
this topic by investigating whether task board customization is beneficial or 
not.

\subsection{Viewers' Consideration of Software Development Artifacts}
There are already several researchers who used eye tracking to investigate a viewer's consideration of a respective software development artifact.

Ahrens et al. \cite{Ahrens.2016} conducted an eye tracking study to analyze how software specifications are read. They identified similar patterns between paper- and screen-based reading. The results contribute awareness by considering readers' interests based on how they use a specification.
Gross and Doerr \cite{Gross.2012} performed an explorative eye tracking study 
to investigate software architects' information needs and expectations from a 
requirements specification. The results allow first insights into the relevance 
of certain artifact types and their notational representations.
Gross et al. \cite{Gross.2012b} extended their previously mentioned eye 
tracking study by analyzing information needs and expectations of usability 
experts. Based on the findings, the authors introduced the idea of a view-based 
requirements specification to fulfill needs of different roles in software 
development.
Santos et al. \cite{Santos.2016} evaluated the effect of layout guidelines for $i^{*}$ goal models on novice stakeholders' ability to understand and review such models. They identified no statistically significant differences in success, time taken or perceived complexity between tasks conducted with well and badly designed model layouts.
Ali et al. \cite{Ali.2012} applied eye tracking to the verification of 
requirements traceability links. Their data analysis allowed the identification 
and ranking of developers' preferred source code entities. Thus, the authors 
defined two weighting schemes to recover traceability links combined with 
information retrieval techniques.

All previous studies apply eye tracking to analyze how specific software 
development artifacts are read by persons with different functions. We follow 
this approach by using eye tracking to investigate the work with a task board. 
Our study specifically focuses on the impact of different task board 
customizations on a task board's usage by team externals respectively new team 
members.

\section{Task Board: Structure and Content}
\label{sec:task-board}
The task board's origins are the \textit{informative workspace} practice of Beck and Andres \cite{Beck.2007} as well as the concept of \textit{information radiator} by Cockburn \cite{Cockburn.2009}. They present first ideas of story cards pinned on a wall or whiteboard. In their books, they offer possible implementations of these concepts.

Cohn \cite{Cohn.2012} describes a first concrete task board design in his book 
``\textit{Agile Estimating and Planning}''. According to his definition, a task 
board consists of up to seven columns to track and visualize a team's progress 
in development. The seven columns are:
\begin{enumerate}
	\item \textit{Stories}: A backlog of all story cards
	\item \textit{To Do}: All task cards to implement particular story cards
	\item \textit{Tests Ready}: Status of a story cards' acceptance tests
	\item \textit{In Process}: Task cards developers have signed up for
	\item \textit{To Verify}: Implemented task cards that need to be verified
	\item \textit{Hours}: Total working hours remaining for particular story 
	cards
	\item \textit{Done}: All implemented and verified task cards
\end{enumerate}
Furthermore, Cohn \cite{Cohn.2012} defines that a task board includes one row for each story card. Each row contains all task cards that are related to the corresponding story card. According to Cohn \cite{Cohn.2012}, the columns \textit{Tests Ready}, \textit{To Verify}, \textit{Hours} and \textit{Done} are optional.

\subsection{Task Board Customizations}
Based on the previously mentioned findings in literature, we considered further 
research papers about the design and content of task boards. Additionally, we 
analyzed different task boards with respect to their design in online galleries 
of team spaces \cite{XP123, Infoq, Scissor}. Thus, we identified three major 
kinds of customization: \textit{modified structures}, \textit{individual 
colored codings}, and \textit{changed card sizes}.

\textbf{Modified structures} are changes regarding the amount and usage of a task board's rows and columns.
Petre et al. \cite{Petre.2012} describe a task board as a vertical surface for 
story cards. This task board has a codified structure to indicate a story 
card's status. Other researchers \cite{PriesHeje.2011, Rubart.2009, Perry.2008} 
report in greater detail about this codified structure. Pries-Heje and 
Pries-Heje \cite{PriesHeje.2011} focus on a task board for Scrum, which 
consists of the four columns \textit{Backlog}, \textit{Task in Progress}, 
\textit{Done}, and \textit{Done Done} corresponding to their description. A 
similar task board structure is mentioned by Rubart and Freykamp 
\cite{Rubart.2009}. The columns of this task board are named \textit{Selected 
Product Backlog}, \textit{Tasks To Do}, \textit{Work In Progress}, and 
\textit{Done}. Perry \cite{Perry.2008} also reports that a simple task board 
has four columns called \textit{Story}, \textit{To Do}, \textit{In Progress}, 
and \textit{Complete}.

All descriptions have in common that the task board structure consists of the 
same four columns with only slightly different labels. However, none of these 
researchers mentions the use of rows on a task board. We could identify two 
variants for the use of rows based on our consideration of team spaces in 
online galleries. The first variant uses one row for each story card, which 
corresponds to Cohn's definition \cite{Cohn.2012}. The second one uses rows 
in specific columns like \textit{To Do} and \textit{Work In Progress} to 
visualize the assignment of developers to story cards. The comparison of these 
insights with Cohn's original task board structure \cite{Cohn.2012} shows clear 
differences regarding the amount and use of a task board's rows and columns 
between theory and practice.

\textbf{Individual colored codings} are colored cards and markers with 
arbitrary meaning which need to be memorized.
Several researchers report the widespread individual use of colored codings on 
task boards. Katsma et al. \cite{Katsma.2013} describe the use of different 
colored cards to indicate various card types, e.g. red for bugs cards. Liskin 
et al. \cite{Liskin.2014} mention colored markers on cards to represent 
assigned developers. Sharp et al. \cite{Sharp.2006, Sharp.2008, Sharp.2009} 
observe the use of colored markers and cards as status indicators and card 
types in four mature XP teams. These findings correspond to our observations of 
the task boards presented in the online galleries. Even though we cannot 
clarify the exact meanings of the used colored codings, we observe that their 
use is widely scattered.

\textbf{Changed card sizes} consider the size of story cards which are used to write down user stories and display them on the task board.
The size of story cards has a wide range. Azizyan et al. \cite{Azizyan.2011} as 
well as Katsma et al. \cite{Katsma.2013} report about story cards the size of 
sticky notes or post-its. In contrast, Perry \cite{Perry.2008} and Sharp et al. 
\cite{Sharp.2009} state that a story card's size can be up to an index card of 
$5 \times 7$ inches.
These insights coincide with our observations of the online galleries. We identified the same range of card sizes from post-its up to index cards.

\subsection{Task Board Designs}
In consideration of the previously described findings, we developed four task 
board designs for our eye tracking study. These designs are based on a dataset 
of real story cards from a completed software project. While one task board 
design is similar to Cohn's initial definition of a task board design 
\cite{Cohn.2012}, each of the other designs takes one of the three major 
customizations into account.

\begin{figure}[!b]
	\begin{subfigure}{.5\textwidth}
		\centering
		\includegraphics[width=\textwidth]{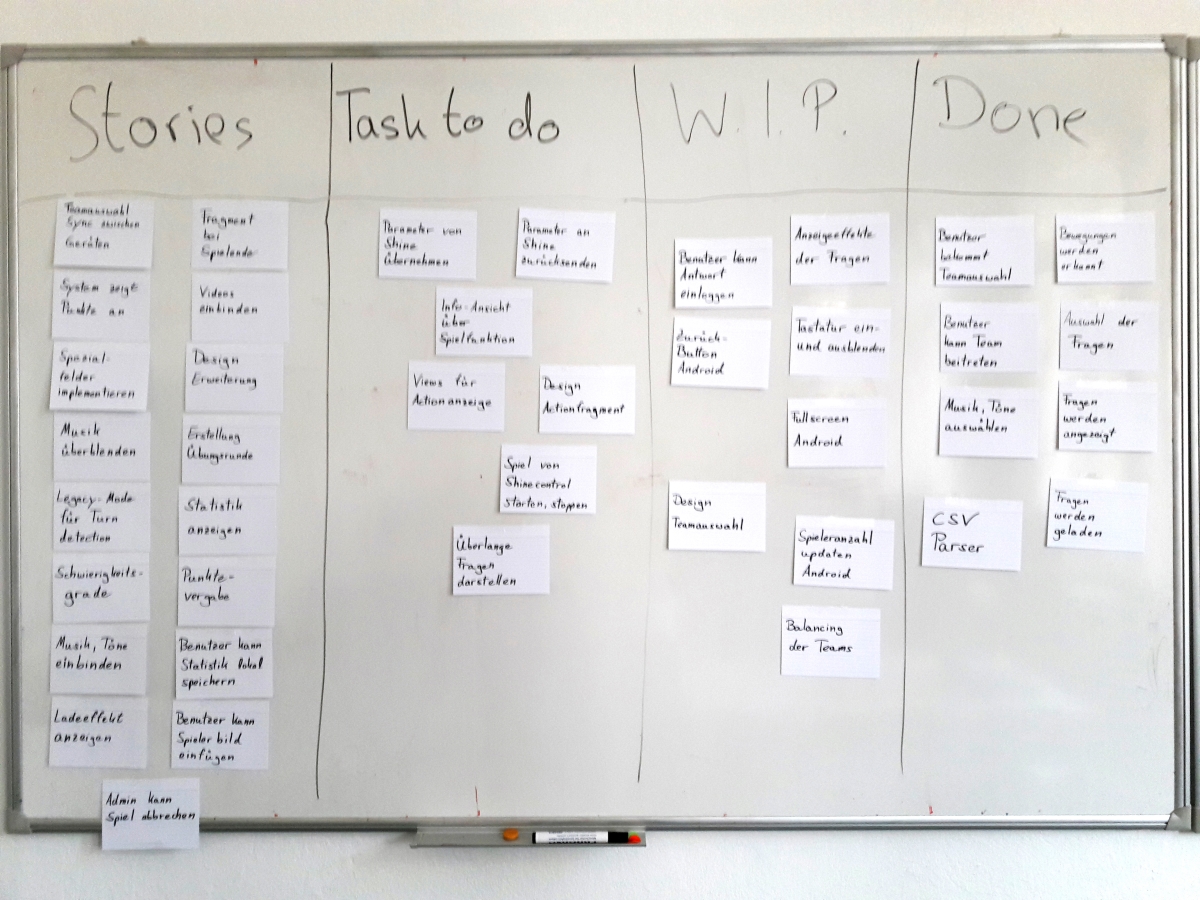}
		\caption{Task board: Original design}
		\label{fig:sfig1}
	\end{subfigure}
	\begin{subfigure}{.5\textwidth}
		\centering
		\includegraphics[width=\textwidth]{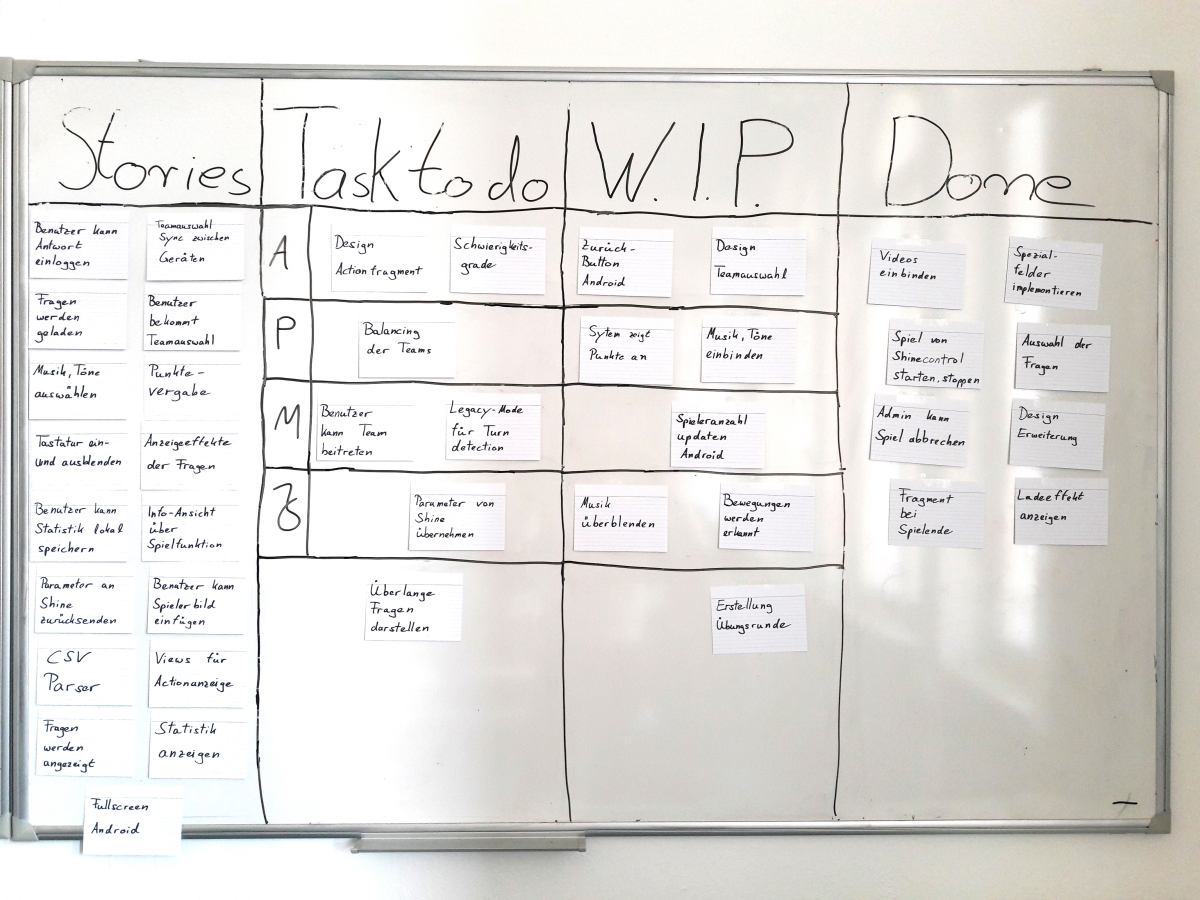}
		\caption{Task board: Modified structures}
		\label{fig:sfig2}
	\end{subfigure}
	\begin{subfigure}{.5\textwidth}
		\centering
		\includegraphics[width=\textwidth]{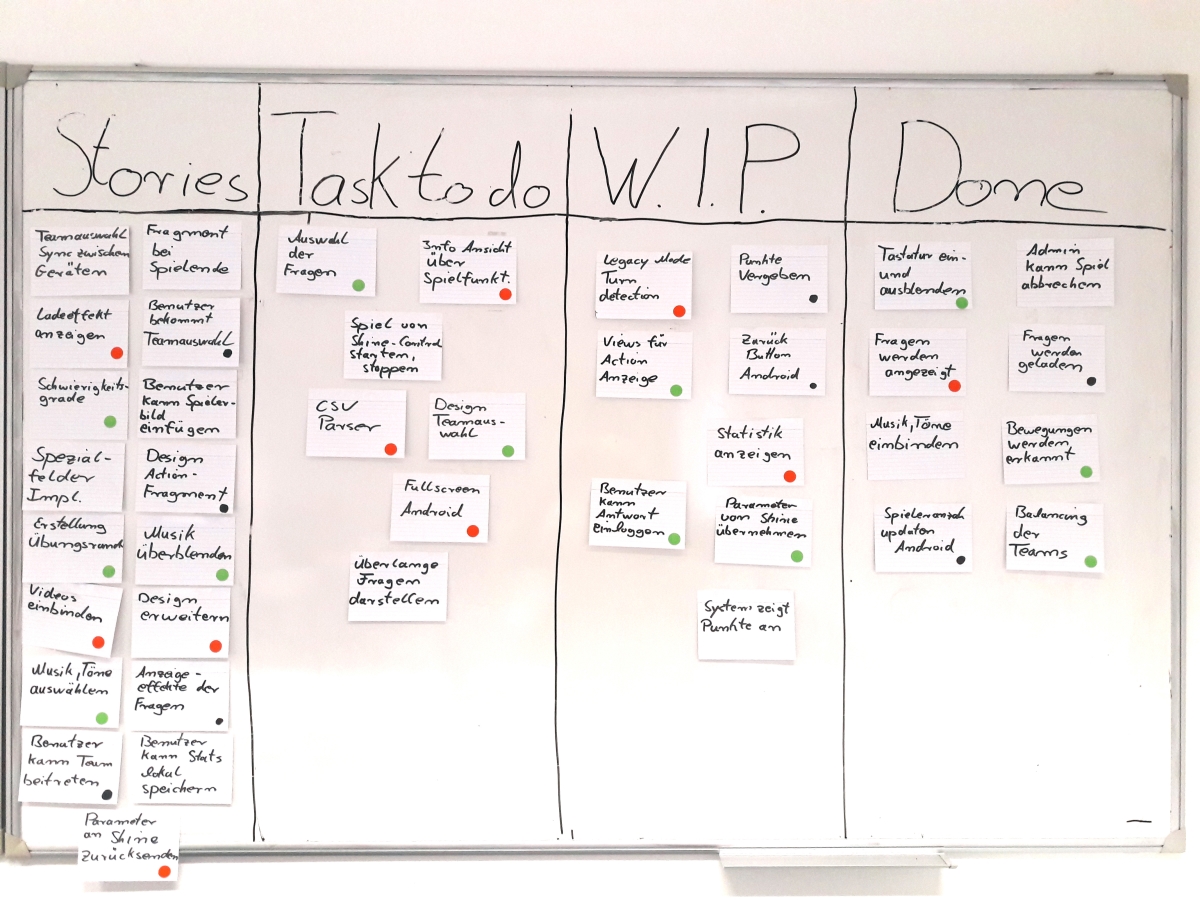}
		\caption{Task board: Individual colored codings}
		\label{fig:sfig3}
	\end{subfigure}
	\begin{subfigure}{.5\textwidth}
		\centering
		\includegraphics[width=\textwidth]{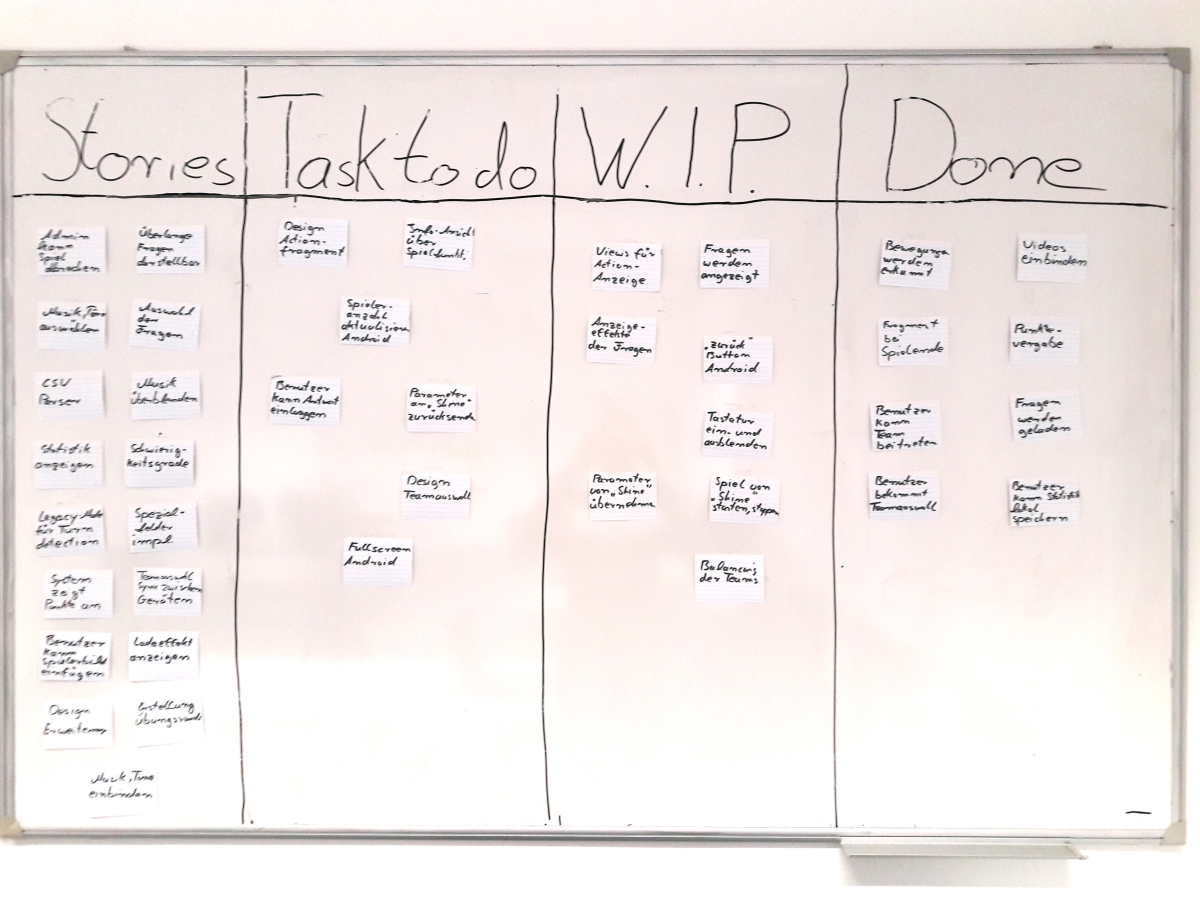}
		\caption{Task board: Changed card sizes}
		\label{fig:sfig4}
	\end{subfigure}
	\caption{Task board designs}
	\label{fig:fig}
\end{figure}

During the design development, we took into account that all task boards 
represent the same content, except for exactly one specific difference 
according to the customizations. \figurename{ \ref{fig:fig}} represents an 
overview of our four task board designs. All task boards have four columns, 
labeled with \textit{Stories}, \textit{Task To Do}, \textit{W.I.P} (Work In 
Progress), and \textit{Done}. These labels are adopted from the original task 
board of the completed software project whose story cards were used. We decided 
to change as little as possible from the original dataset. Therefore, we retain 
the labels of the task board since they are similar to the previously mentioned 
ones. Furthermore, these four columns cover all three obligatory columns 
corresponding to Cohn's definition \cite{Cohn.2012}.

\figurename{ \ref{fig:sfig1}} presents the task board with an original design which is similar to Cohn's definition \cite{Cohn.2012}. This task board does not have Cohn's row structure \cite{Cohn.2012} since the used dataset of real story cards did not consider this aspect. Therefore, the story cards could not be grouped to achieve a reasonable row structure.

\figurename{ \ref{fig:sfig2}} shows the task board with modified structures. We 
decided to use the second variant of additional rows over specific columns 
since Cohn's row structure \cite{Cohn.2012} was not applicable to the used 
dataset. We did not add additional columns to change only one structural 
aspect. Thus, we added rows over the columns \textit{Task To Do} and 
\textit{W.I.P.} to visualize the assignment of developers to story cards. Each 
row starts with a letter that represents one developer.

\figurename{ \ref{fig:sfig3}} represents the task board with individual colored 
codings. In accordance with literature \cite{Liskin.2014, Sharp.2006}, we added 
colored markers on the right lower corner of the story cards. Each of the three 
colors (green, orange and blue) represents one developer and his assignment to 
the corresponding story card.

\figurename{ \ref{fig:sfig4}} illustrates the task board with changed card sizes. We decided to minimize the story cards to sticky note size (ca. $4 \times 6$ inches), since story cards have originally index card size (ca. $5 \times 7$ inches).

All task boards have the same amount of handwritten story cards whose content 
is based on the real dataset. The first three task boards (see \figurename{ 
\ref{fig:sfig1}}, \figurename{ \ref{fig:sfig2}}, and \figurename{ 
\ref{fig:sfig3}}) contain 40 story cards of index card size. The last task 
board (see \figurename{ \ref{fig:sfig4}}) contains 40 story cards of post-it 
note size. While the amount and general position of the story cards are the 
same in each column and task board, we shuffled the story cards before placing 
them on the task boards. Thus, we achieved a random placement regarding the 
story cards' content and no task board equals exactly any other.

\section{Eye Tracking Study}
\label{sec:controlled-experiment}
The aim of our eye tracking study was to understand whether task 
board customization facilitates identifying a particular story card faster 
compared to an original task board design.
We proceeded to achieve this aim by comparing the original task board design 
with each of the three task board customizations. Such an investigation enables 
us to judge whether the original task board design or the respective task board 
customization should be preferred. We were interested in answering the 
following research question:

\begin{itemize}[leftmargin=9mm]
	\item[RQ:] Does the respective task board customization facilitate 
	identifying a particular story card faster compared to the original task 
	board design?
\end{itemize}

To answer the research question, we tested the following hypotheses for each of 
the three task board customizations:

\begin{itemize}[leftmargin=8mm]
	\item[$H_{0}$:] There is no speed difference in identifying a particular 
	story card between the original task board design and the respective task 
	board customization.
	\item[$H_{1}$:] There is a speed difference in identifying a particular 
	story card between the original task board design and the respective task 
	board customization.
\end{itemize}

\subsection{Study Design}
In this study, we performed three separate within-subjects experiments with 
counterbalancing. The dependent variable was the task completion time for 
identifying a particular story card. The independent variable was the task 
board design with two levels: the original task board design and one of the 
three task board customizations. We measured the task completion time by 
observing the participants with the \textit{SMI Eye Tracking 
Glasses}\footnote{https://www.smivision.com/eye-tracking/product/eye-tracking-glasses/}.
 Each experiment represents a scenario in which the participant joins an 
ongoing development project as a new team member who has to work with the 
existing task board. We decided to focus on the perspective of a new team 
member since a task board should support a fast and easy project overview for 
everyone, i.e. the team and team externals respectively new team members. If a 
new team member already benefits from a customization, a whole team should also 
benefit from it.

We analyzed task completion times with a two-tailed paired samples $t$-test 
at a significance level of $p = 0.05$. This allows us to determine 
whether the respective task board customization leads to a statistically 
significant speed difference in identifying a particular story card compared 
to the original task board design. Thus, we can identify whether a particular 
task board customization is beneficial for a task board's usage. An existing 
speed difference would allow us to reject $H_{0}$, while a missing one would 
not allow such a rejection.

\subsection{Study Procedure}
The eye tracking study was carried out with $30$ participants consisting of 
$10$ undergraduate and $20$ graduate students of computer science. All 
participants had basic knowledge about agile software development and were 
close to the next step in graduation. Thus, they represent potential new team 
members in a software development team, which corresponds to our target 
population.

All in all, the whole eye tracking study with all three experiments was carried 
out within three months. Each experiment compared the original task board 
design with one of the three major task board customizations. We randomly 
assigned each participant to one of the three experiments. In each experiment, 
we conducted $10$ separate sessions each with one of the $10$ assigned 
participants. Each session included an introduction to the experiment with its 
task of considering two task boards. In this context, we explained the basic 
concept of a task board. Depending on the experiment, we assigned the letter 
``J'' (see \figurename{ \ref{fig:sfig2}}) respectively the color ``green'' (see 
\figurename{ \ref{fig:sfig3}}) to the participant since the task boards with 
modified structures respectively individual colored codings required the 
assignment of a row or color to the participant. After the calibration of the 
\textit{SMI Eye Tracking Glasses} for the participant, we captured their 
examination of the task board. We repeated the same process for the second task 
board design.

\subsection{Analysis and Results}
\tablename{ \ref{tb:tct}} shows the measured task completion times of each 
participant for the particular experiment and respective task board design. The 
first five subjects of each experiment (see \tablename{ \ref{tb:tct}}, 
\textit{Group $1$}) received the original task board design first and then the 
customized one. The other five subjects of each experiment (see \tablename{ 
\ref{tb:tct}}, \textit{Group $2$}) received the designs in reversed order. For 
each experiment, we verified that the data is normally distributed by applying 
the \textit{Shapiro-Wilk} test. Subsequently, we performed the two-tailed 
paired samples $t$-tests at a significance level of $p = 0.05$. Thus, we can 
determine whether an observed difference exists due to the test conditions or 
by chance. Additionally, we calculate Cohen's $d$ which is the most common 
type of effect size for $t$-tests that indicates whether or not the difference 
between two groups' mean is large enough to have practical relevance 
independently from statistical significance.

\begin{table}[!b]
	\centering
	\caption{Experiment results -- Task completion time [s]}
	\label{tb:tct}
	\begin{tabular}{|c||c|c|c||c|c|c||c|c|c|}
		\cline{2-10}
		\multicolumn{1}{c|}{\multirow{2}{*}{}} &
		\multirow{2}{*}{\textbf{Subj.}} & 
		\multicolumn{2}{c||}{\textbf{Experiment 1}} & 
		\multirow{2}{*}{\textbf{Subj.}} & 
		\multicolumn{2}{c||}{\textbf{Experiment 2}} & 
		\multirow{2}{*}{\textbf{Subj.}} & 
		\multicolumn{2}{c|}{\textbf{Experiment 3}} \\ \hhline{~~--||~--||~--}
		\multicolumn{1}{c|}{\multirow{2}{*}{}} & & 
		\multicolumn{1}{c|}{\textbf{Original}} & \textbf{Structures} & & 
		\multicolumn{1}{c|}{\textbf{Original}} & \textbf{Codings} & & 
		\multicolumn{1}{c|}{\textbf{Original}} & \textbf{Cards} \\ 
		\hhline{-b|===::===::===|}
		\multirow{5}{*}{\rotatebox[origin=c]{90}{Group 1}} &
		\multicolumn{1}{|c|}{P1} & 16 & 10 & \multicolumn{1}{|c|}{P3} & 16 & 15 
		& \multicolumn{1}{|c|}{P4} & 10 & 11 \\ 
		& \multicolumn{1}{|c|}{P2} & 18 & 4 & \multicolumn{1}{|c|}{P14} & 16 & 
		22 & \multicolumn{1}{|c|}{P5} & 13 & 27 \\
		& \multicolumn{1}{|c|}{P11} & 16 & 11 & \multicolumn{1}{|c|}{P15} & 11 
		& 12 & \multicolumn{1}{|c|}{P24} & 30 & 36 \\ 
		& \multicolumn{1}{|c|}{P12} & 12 & 9 & \multicolumn{1}{|c|}{P16} & 10 & 
		9 & \multicolumn{1}{|c|}{P25} & 4 & 19 \\ 
		& \multicolumn{1}{|c|}{P13} & 19 & 4 & \multicolumn{1}{|c|}{P17} & 13 & 
		16 & \multicolumn{1}{|c|}{P26} & 19 & 18 \\ \hhline{|=::===::===::===|}
		\multirow{5}{*}{\rotatebox[origin=c]{90}{Group 2}} & 
		\multicolumn{1}{|c|}{P6} & 18 & 10 & \multicolumn{1}{|c|}{P18} & 12 & 
		20 & \multicolumn{1}{|c|}{P23} & 22 & 12 \\ 
		& \multicolumn{1}{|c|}{P7} & 8 & 16 & \multicolumn{1}{|c|}{P19} & 13 & 
		16 & \multicolumn{1}{|c|}{P27} & 9 & 19 \\ 
		& \multicolumn{1}{|c|}{P8} & 12 & 9 & \multicolumn{1}{|c|}{P20} & 11 & 
		13 & \multicolumn{1}{|c|}{P28} & 19 & 28 \\ 
		& \multicolumn{1}{|c|}{P9} & 13 & 15 & \multicolumn{1}{|c|}{P21} & 5 & 
		6 & \multicolumn{1}{|c|}{P29} & 17 & 27 \\ 
		& \multicolumn{1}{|c|}{P10} & 18 & 10 & \multicolumn{1}{|c|}{P22} & 10 
		& 14 & \multicolumn{1}{|c|}{P30} & 12 & 16 \\ 
		\cline{1-1} \hhline{~|===::===::===|}
		\multicolumn{1}{c|}{} & \multicolumn{1}{|c|}{Mean} & 15.0 & 9.8 & 
		\multicolumn{1}{|c|}{Mean} 
		& 11.7 & 14.3 & \multicolumn{1}{|c|}{Mean} & 15.5 & 21.3 \\ 
		\hhline{~---||---||---}
		\multicolumn{1}{c|}{} & \multicolumn{1}{|c|}{SD} & 3.6 & 3.9 & 
		\multicolumn{1}{|c|}{SD} & 3.2 
		& 4.7 & \multicolumn{1}{|c|}{SD} & 7.5 & 7.9 \\ \cline{2-10}
	\end{tabular}
\end{table}

\begin{table}[!b]
	\centering
	\caption{Two-tailed paired samples $t$-test}
	\label{tb:t-test}
	\begin{tabular}{|c||c||c||c|}
		\hline
		\textbf{Experiment} & \textbf{1} & \textbf{2} & \textbf{3} \\ 
		\hhline{|=::=::=::=|}
		\textbf{ Calculated $t\textnormal{-}value$ } & -2.39 & 2.86 & 2.41 \\ 
		\hhline{|-||-|b|-|b|-|}
		\textbf{\begin{tabular}[c]{@{}c@{}}$t\textnormal{-}value$ from table\\ 
		$(df = 9, \alpha = 0.05)$\end{tabular}} & \multicolumn{3}{c|}{2.26} \\ 
		\hhline{|-||-b-b-|}
		\textbf{ Calculated $p\textnormal{-}value$ } & 0.04 & 0.02 & 0.04 \\ 
		\hhline{|-||-||-||-|}
		\textbf{Result} & \multirow{2}{*}{ Significant } & \multirow{2}{*}{ Significant } & \multirow{2}{*}{ Significant } \\
		\textbf{$(p\textnormal{-}value \leq 0.05 ?)$} & & & \\ 
		\hhline{|-||-||-||-|}
		\textbf{Cohen's $d$} & 0.76 & 0.90 & 0.76 \\ \hline
	\end{tabular}
\end{table}

In \tablename{ \ref{tb:t-test}}, we present the results of our conducted 
two-tailed paired samples $t$-tests and their effect size $d$.

The analysis of the first experiment yields a significant difference in the 
task completion times for the original task board design ($M = 15.0 s, SD = 3.6 
s$) and the modified structures ($M = 9.8 s, SD = 3.9 s$); $t(9) = -2.39, p = 
0.04$. Hence, $H_{0}$ can be rejected for the first experiment. Modified 
structures shorten time to identify a particular story card compared to the 
original task board design. The value of Cohen's $d$ is $0.76$, which is close 
to the threshold of $0.8$ for a large effect \cite{Cohen.1992}. Hence, the 
identified difference has almost large practical relevance.

The $t$-test of the second experiment shows a significant difference between 
the task completion times for the original task board design ($M = 11.7 s, SD = 
3.2 s$) and the individual colored codings ($M = 14.3 s, SD = 4.7 s$); $t(9) = 
2.86, p = 0.02$. The null hypothesis $H_{0}$ can be rejected for the second 
experiment. Consequently, the original task board design allows to identify 
a particular story card faster compared to the individual colored codings. 
Cohen's $d$ is $0.90$ and thus greater than the threshold of $0.8$ for a large 
effect \cite{Cohen.1992}. The determined difference between the individual 
colored codings and the original task board design has large practical 
relevance.

The results of the third experiment also show a significant difference in the 
task completion time for the original task board design ($M = 15.5s, SD = 
7.5s$) and the changed card sizes ($M = 21.3 s, SD = 7.9 s$); $t(9) = 2.41, p = 
0.04$. Consequently, we can reject $H_{0}$. This leads to the insight that 
changed card sizes increase the required time for identifying a particular 
story card compared to the original task board. The calculated effect size $d$ 
is $0.76$ and thus close to the threshold of $0.8$. We identified a difference 
between changed card sizes and the original task board design that has 
almost large practical relevance.

\subsection{Interpretation}
Our findings provide insights with respect to the influence of task board 
customizations in comparison with an original task board design. Whereas 
modified structures shorten time to identify a particular story card, 
individual colored codings and changed card sizes increase the required time.

The performed $t$-tests substantiate that there is a statistically significant 
difference between the respective task board customization and the original 
task board design. Our results indicate that customizing a task board's 
structure supports its usage. In case of customization, agile teams should 
focus on adjusting the structure of a task board according to their needs. 
Since this customization supports the work of new team members who are 
unfamiliar with the task board, we assume that a whole team will also benefit 
from it. Such a customized task board provides a fast and easy project 
overview for everyone, i.e. the team and team externals respectively 
new team members. Thus, the task board complies more precisely with the 
agile practice \textit{informative workspace}.

However, according to our results, not every customization is beneficial for a 
task board's usage. Adjustments on story cards such as individual colored 
codings or changed card sizes lead to an increased amount of time to identify a 
particular story card. Even though these two kinds of customization do not 
necessarily support a task board's usage, they are extensively applied in 
practice by agile teams \cite{Liskin.2014, Sharp.2006, Sharp.2008, Sharp.2009}. 
Therefore, our findings are in conflict with the observed widely distributed 
use of these customizations.

In total, we identified a statistically significant difference in each of the 
three experiments. Each difference indicates that one of the two compared task 
board designs (customized vs. original) allows identifying a particular story 
card faster. All findings have an almost large effect size $d$ that emphasizes 
their practical relevance. According to our results, modified structures should 
be preferred compared to the original task board design, which is, in turn, 
preferable to individual colored codings and changed card sizes. Hence, the 
original task board design is a good solution. In case of customization, 
however, agile teams may be better served by adjusting their task board's 
structure instead of its story cards. As an answer to our research question, we 
can summarize:
\begin{itemize}
	\item[A:] We identified that only the modified structures allow identifying 
	a particular story card faster compared to the original task board design. 
	Both of the other customizations result in an increased amount of time. 
	Hence, adjusting a task board's structure is the only beneficial option of 
	all investigated customizations.
\end{itemize}

\subsection{Threats to Validity}
In the presented eye tracking study, we considered threats to construct, 
external, internal and conclusion validity corresponding to Wohlin et al. 
\cite{Wohlin.2012}.

\textbf{Construct validity:}
We selected the content for the story cards from a completed software project. 
All task boards (see \figurename{ \ref{fig:fig}}) were based on this content. 
Thus, we have a mono-operation bias since we only use one dataset for the task 
boards' content. As a consequence, the constructed task boards do not convey a 
comprehensive overview of the task boards' complexity in practice. However, we 
expected that the amount of $40$ handwritten story cards and their different 
arrangement on each task board result in sufficient realistic complexity for 
the participants.
Another threat to validity was the participants falsely reporting having 
finished. Our experiments required the exact measuring of the task 
completion time. However, people are afraid of being evaluated and they are 
inclined to convey the impression of being better than they really are. 
Therefore, this human tendency endangered the outcome of our experiment. We 
counteracted this threat by using eye tracking combined with an additional 
acoustical statement of the participants when they identified the particular 
story card. Thus, we could determine the exact task completion time of each 
participant beyond doubt.
The single use of eye tracking is a further threat to validity. This 
mono-method bias is problematic since it only allows a restricted explanation 
of our findings. However, we focused on an objective measure instead of a 
subjective one since objective measures can be reproduced more easily and are 
thus more reliable.
The given task of identifying a particular story card caused an interaction of 
testing and treatment. The comparison of task boards with the given task could 
imply to find the story card as fast as possible. Even though we did not 
mention to measure task completion time, the participants could be aware of the 
time as a factor. Instead of understanding the task board designs, they could 
only have tried to be as fast as possible. We mitigated this threat to validity 
by using eye tracking. Thus, we could observe how the participants examine the 
task boards and make sure that all of them took the respective design into 
account.

\textbf{External validity:}
The choice of involving almost graduated students as participants, and the use 
of data from a completed software project produced a good level of realism. At 
the same time, the experimental setting endangered the external validity since 
the environment was different from the real world. None of the task boards had 
true pragmatic value for the participants since none of them had a genuine 
working task with the task board. Future evaluation should be done on real 
industry projects with team members that truly work with the task board.

\textbf{Internal validity:}
In our eye tracking study, the three experiments were distributed over three 
months altogether. This large period of time could have an effect on the 
participants' motivation to contribute to our study. However, we could not 
compare all task board designs within one experiment due to the use of eye 
tracking, which is time-consuming as well as exhausting for the participants. A 
single session with one participant required as much as $25$ minutes for the 
comparison of two task board designs. Additionally, we could mitigate possible 
learning effects since all task board designs equaled one another except for 
exactly one specific difference with respect to the corresponding customization.

\textbf{Conclusion validity:}
We decided to use eye tracking to improve the reliability of our results since an objective measuring is easier to reproduce and it is more reliable than a subjective one.
Additionally, we only selected students as participants who were close to their 
graduation. Hence, they form a more homogeneous group which counteracts the 
threat of erroneous conclusions. Therefore, we mitigated the risk that the 
variation due to the subjects' random heterogeneity is larger than due to 
the investigated task board designs.

\section{Discussion}
\label{sec:discussion}
This presented work investigates the \textit{task board} as one implementation 
of the agile practice \textit{informative workspace} and the benefit of task 
board customization.

Although agile teams use task boards in a similar manner, they tend to 
customize their task boards according to their needs. Combined customizations 
such as modified structures, individual colored codings, and changed card sizes 
lead to complexity, which impedes a task board's maintenance and 
comprehensibility. The increased effort is in conflict with a task board's 
underlying agile practice of fast and easy project overview for everyone, i.e. 
the team and team externals respectively new team members.
We performed an eye tracking study to analyze whether there is a significant 
speed difference in the time required to identify a particular story card 
between the original task board design and the respective task board 
customization.

We contribute the insight that only modified structures improve a task board's 
usage. In contrast, individual colored codings and changed card sizes did not 
improve performance beyond the original design.

The modified structures are the only beneficial customization. We assume that 
the additional rows improve the arrangement of the story cards. Spatially close 
object seems to be grouped since they are perceived as belonging to each other. 
This effect is called \textit{law of proximity}, which is part the 
\textit{Gestalt Principles} \cite{Palmer.1999}. 
The additional rows influence the story card's visual appearance by position 
without further support. The story cards' improved proximity simplifies a 
viewer's consideration of the task board. This finding can help agile teams to 
rethink their task board in order to improve it. They may be better served by 
focusing on the original task board design and by only adjusting its structure 
according to their needs. Thus, they can create a task board which complies 
more precisely with the concept of fast and easy project overview for everyone.

In contrast, individual colored codings and changed card sizes are not 
beneficial. The missing benefit of individual colored codings is caused by a 
counteracting effect of combined laws of the \textit{Gestalt Principles}. 
According to the \textit{law of similarity}, using colors for similar objects 
supports the visual appearance of belonging together. At the same time, the 
story cards' spatial arrangement complies with the \textit{law of proximity}. 
The colored markers are more difficult to perceive since the \textit{law of 
proximity} dominates the \textit{law of similarity}. Therefore, individual 
colored codings do not provide a benefit for customizing a task board. The 
changed card sizes are not an improving task board customization, either. 
According to our results, a viewer's effort increases by considering and 
recognizing smaller story cards to identify a particular one. Smaller story 
cards are more difficult to perceive and read, which complicates a task board's 
clarity. Thus, changed card sizes provide no benefit, either.

The impact of the different task board designs on the performance of a 
single team is low. Even if a team member identifies $120$ times a day a story 
card with an average saving of $5$ seconds per identification, his total saving 
would only be $10$ minutes per workday.
The benefit of our results is the finding that the original design of a task 
board by Cohn \cite{Cohn.2012} with its underlying agile practices constitutes 
already a good solution for a single team to be productive. Even though agile 
approaches offer corresponding degrees of freedom for customization, in the 
worst case each team of a company has its own specific task board design. Due 
to the wide variety of customization options, the individual task boards 
complicate the collaboration across teams and the work of team externals. Thus, 
the collaboration performance of multiple teams, as well as the work of team 
externals, can be improved by focusing on one consistent and beneficial task 
board design.

All in all, we can conclude that not each kind of task board customization is 
beneficial. Based on our findings, we agree with Berczuk \cite{Berczuk.2007}: 
Teams are better served by adjusting their task boards as little as possible.
As a consequence, agile teams should rethink their current task board design 
with respect to the applied customizations. The original task board design (see 
\figurename{ \ref{fig:sfig1}}) is already a good solution. However, if 
customization is desired, teams should focus on adjusting a task board's 
structure since only this kind of customization improves the use of a 
task board according to our results.

\section{Conclusion}
\label{sec:conclusion}
This work contributes the insight that not every kind of task board 
customization is beneficial. Agile teams tend to extensively customize their 
task boards according to their needs \cite{Sharp.2009}. However, the use of 
modified structures, individual colored codings, and changed card sizes 
impede work with a task board. Thus, task board customization is in 
conflict with the agile practice of fast and easy project overview for 
everyone, i.e. the team and team externals.

We performed an eye tracking study consisting of three separate experiments 
comparing an original task board design with each of three identified major 
task board customizations. Based on these results, we identified statistically 
significant differences in all three experiments. These findings show that 
modified structures such as additional rows support a task board's usage with 
respect to the used exemplary main task of identifying a particular story card. 
In contrast, individual colored codings and changed card sizes do not improve 
performance beyond the original design.

Our work points to the conclusion that agile teams should rethink their current 
task board design. They may be better served by focusing on the original task 
board design and applying carefully selected adjustments. In case of 
customization, teams should adjust the task board's structure since this is the 
only beneficial kind of customization. Additionally, such a customized task 
board design complies more precisely with its implemented agile practice.

\section*{Acknowledgment}
This work was supported by the German Research Foundation (DFG) under ViViReq 
($2017$ -- $2019$). We follow ethical guidelines of the Central Ethics 
Commission of our university. They regulate subject information and rights. 
Since recognizable persons should not be visible on distributed video, our data 
is archived internally for future reference.

\bibliographystyle{splncs03}
\bibliography{refs}

\end{document}